\documentclass{mn2e}

\usepackage{psfig}

\newcommand\beq{\begin{equation}}
\newcommand\eeq{\end{equation}}

\begin{document}

\setlength\topmargin{-3.3pc}
 
\title[How rapidly do neutron stars spin at birth?]{How rapidly do
neutron stars spin at birth? Constraints from archival X-ray observations
of extragalactic supernovae}

\author[Perna et al.]
{Rosalba Perna$^{1}$, 
Roberto Soria$^{2}$,
Dave Pooley$^{3}$,
Luigi Stella$^{4}$\\
$^1$ {\sl JILA and Department of Astrophysical and Planetary Sciences, University
of Colorado, Boulder, CO, 80309}\\
$^2$ {\sl MSSL, University College London, Holmbury St mary, Dorking RH5 6NT, UK }\\
$^3$ {\sl Astronomy Department, University of Wisconsin-Madison
475 North Charter st., Madison, WI 53706, USA }\\
$^4$ {\sl INAF - Osservatorio Astronomico di Roma, Via Frascati 33, I-00040 Rome, Italy}\\}

\maketitle

\begin{abstract}

Traditionally, studies aimed at inferring the distribution of birth
periods of neutron stars are based on radio surveys.  Here we propose
an independent method to constrain the pulsar spin periods at birth
based on their X-ray luminosities.  In particular, the observed
luminosity distribution of supernovae poses a constraint on the
initial rotational energy of the embedded pulsars, via the
$L_X-\dot{E}_{\rm rot}$ correlation found for radio pulsars, and under
the assumption that this relation continues to hold beyond the
observed range.  We have extracted X-ray luminosities (or limits) for
a large sample of historical SNe observed with {\em Chandra}, {\em
XMM} and {\em Swift}, that have been firmly classified as
core-collapse supernovae. We have then compared these observational
limits with the results of Monte Carlo simulations of the pulsar X-ray
luminosity distribution, for a range of values of the birth
parameters. We find that a pulsar population dominated by millisecond
periods at birth is ruled out by the data.

\end{abstract}

\section{Introduction}

Modeling the observed properties of the Galactic population of radio
pulsars, with the purpose of inferring their intrinsic properties, has
been the subject of extensive investigation for several decades
(e.g. Gunn \& Ostriker 1970; Phinney \& Blandford 1981; Lyne et
al. 1985; Stollman 1987; Emmering \& Chevalier 1989; Narayan \&
Ostriker 1990; Lorimer et al. 1993; Hartman et al. 1997; Cordes \&
Chernoff 1998; Arzoumanian, Cordes \& Chernoff 2002; Vranesevic et al
2004; Faucher-Giguere \& Kaspi 2006; Ferrario \& Wickramasinghe 2006).
Since the fraction of pulsars that can be detected close to their
birth constitutes a negligible fraction of the total sample, these
studies generally use the {\em present day} observed properties of
pulsars (namely their period $P$ and period derivative $\dot{P}$),
together with some assumptions about their time evolution, to
reconstruct the birth distribution of periods and magnetic fields for
the pulsar population. These analyses also need to make
assumptions about pulsar properties and their evolution (such as, for example,
the exact shape of the radio beam and its dependence on the period),
as well as overcome a number of selection effects.  Results from
various investigations have often been conflicting, with some studies
favoring initial periods in the millisecond range (e.g. Arzoumanian
et al. 2002), and others instead finding more likely periods in the
range of several tens to several hundreds of milliseconds
(e.g. Faucher-Giguere \& Kaspi 2006). The efforts put over the
years into this area of research stem from the fact that the birth
properties of neutron stars (NSs) are intimately related to the
physical processes occurring during the supernova (SN) explosion and in the
proto-neutron star. As such, they bear crucial information on the
physics of core-collapse SNe, in which most are thought to be formed.  

Besides the inferences on the birth parameters from the radio
population discussed above, we show here that constraints can be
derived also from the X-rays.  Young, fast rotating neutron stars are
indeed expected to be very bright in the X-rays. In fact,
observationally there appears to be a correlation between the
rotational energy loss of the star, $\dot{E}_{\rm rot}$, and its X-ray
luminosity, $L_x$. This correlation was noticed by Verbunt et
al. (1996), Becker \& Trumper (1997), Seward \& Wang (1988), Saito
(1998) for a small sample of objects, and later studied by Possenti et
al. (2002; P02 in the following) for the largest sample of pulsars
known to date.

Combining the birth parameters derived from the radio (which determine
the birth distribution of $\dot{E}_{\rm rot}$ for the pulsars), with
the empirical $L_x - \dot{E}_{\rm rot}$ correlation, the distribution
of X-ray luminosity can be predicted for a sample of pulsars with a
certain age distribution.  The above calculation was performed by
Perna \& Stella (2004). They found that the birth parameters derived
by Arzoumanian et al. (2002), together with the $L_x - \dot{E}_{\rm
rot}$ correlation derived by P02, yield a sizable fraction of sources
with luminosities $\ga 10^{39}$ erg/s, which could hence constitute
potential contributors to the observed population of ultra luminous
X-ray sources (ULXs) observed in nearby galaxies (e.g. Fabbiano \&
White 2003; Ptak \& Colbert 2004). Obviously, these predictions were
heavily dependent on the assumed initial birth parameters (the periods
especially) of the pulsar population.

In this paper, we propose a new, independent method to constrain the
pulsar spin periods at birth from X-ray observations, and hence also
assess the contribution of young, fast rotating NSs to the population
of bright X-ray sources.  Since neutron stars are born in supernova
explosions, and very young pulsars are still embedded in their
supernovae, the X-ray luminosity of the SNe provides an upper limit to
the luminosity of the embedded pulsars.  We have analyzed an extensive
sample of historical SNe whose position has been observed by {\em
Chandra}, {\em XMM} or {\em Swift}, and studied their X-ray
counterparts. We measured their X-ray luminosities, or derived a limit
on them in the cases of no detection.  A comparison between these
limits and the theoretical predictions for the distribution of pulsar
X-ray luminosities shows that, if the assumed initial spins are in the
millisecond range, the predicted distribution of pulsar X-ray
luminosities via the $L_x - \dot{E}_{\rm rot}$ correlation is highly
inconsistent with the SN data.  Our analysis hence suggests that a
substantial fraction of pulsars cannot be born with millisecond
periods.

The paper is organized as follows: in \S2, we describe the method by
which the SN X-ray flux measurements and limits are extracted, while
in \S3 we describe the theoretical model for the distribution of the
X-ray luminosity of young pulsars.  A comparison between the
theoretical predictions and the data is performed in \S4, while the
results are summarized and discussed in \S5.

\section{X-ray analysis of historical Supernovae observed by {\em Chandra},
{\em XMM} and {\em Swift}}

We compared and combined the CfA List of
Supernovae\footnote{http://cfa-www.harvard.edu/iau/lists/Supernovae.html,
compiled by The Central Bureau for Astronomical Telegrams at the
Harvard-Smithsonian Center for Astrophysics.}, the Padova-Asiago
Catalogue\footnote{http://web.pd.astro.it/supern/snean.txt}, the
Sternberg Catalogue\footnote{VizieR On-line Data Catalog: II/256}
(Tsvetkov et al. 2004), and Michael Richmond's Supernova
Page\footnote{http://stupendous.rit.edu/richmond/sne/sn.list}, to
create a list of unambiguosly identified core-collapse SNe (updated to
2007 April). We cross-correlated the SN positions with the catalogues
of {\it Chandra}/ACIS, {\it XMM-Newton}/EPIC and {\it Swift}/XRT
observations\footnote{Search form at http://heasarc.nasa.gov}, to
determine which SN fields have been observed by recent X-ray missions
({\it ASCA} was excluded because of its low spatial resolution, and
{\it ROSAT} because of its lack of 2--10 keV sensitivity). For the
{\it Chandra} ACIS-S data, we limited our search to the S3 chip.  We
obtained a list of $\sim 200$ core-collapse SNe whose positions
happened to be in a field observed at least once after the event.
From the list, we then selected for this paper all the core collapse
SNe with unambiguos subtype classification (Type Ib/c, Type IIn, IIL,
and IIP and IIb).  That is about half of the total sample.  We leave
the analysis of the other $\sim 100$ SNe (classified generically as
Type II) to a follow-up paper.

\setcounter{table}{0}
\begin{table*}
\begin{tabular}{lllllll}
SN & Host galaxy & Type & Age (yr) & $L_{2-10\;keV} (\rm erg/s)$ & Instrument & Observation date\\
\hline
1923A & N5236 &IIP&77.3& $<6.0\times 10^{35}$ & ACIS & 2000-04-29, 2001-09-04\\
1926A & N4303 &IIP&75.3& $<1.4\times 10^{37}$ & ACIS & 2001-08-07 \\
1937A & N4157 &IIP&67.3& $<1.3\times 10^{37}$ & EPIC  & 2004-05-16\\
1937F & N3184 &IIP&62.1& $<2.7\times 10^{36}$& ACIS & 2000-01-08, 2000-02-03\\
1940A & N5907 &IIL&63.0& $<1.0 \times 10^{37}$ & EPIC & 2003-02-20, 2003-02-28\\
1940B & N4725 &IIP&62.6& $<8.6\times 10^{36}$ & ACIS & 2002-12-02\\
1941A & N4559 &IIL&60.2& $<5.5\times 10^{36}$& ACIS & 2001-01-14, 2001-06-04, 2002-03-14\\
1948B & N6946 &IIP&55.1& $<4.7\times 10^{35}$& ACIS & 2001-09-07, 2002-11-25, 2004-10-22, 2004-11-06, 2004-12-03\\
1954A & N4214 &Ib & 48.9& $<1.6 \times 10^{35}$& ACIS & 2003-03-09\\
1959D & N7331 &IIL&41.6& $<2.2\times 10^{37}$& ACIS & 2001-01-27\\
1961V & N1058 &IIn&38.3& $<6.1\times 10^{37}$& ACIS & 2000-03-20\\
1962L & N1073 &Ic &41.2& $< 4.7 \times 10^{37}$& ACIS & 2004-02-09\\
1962M & N1313 &IIP&40.3& $<3.7\times 10^{35}$& ACIS & 2002-10-13, 2002-11-09, 2003-10-02, 2004-02-22\\
1965H & N4666 &IIP&37.7& $<1.5\times 10^{38}$ & ACIS & 2003-02-14\\
1965L & N3631 &IIP&37.8& $<5.7\times 10^{36}$& ACIS & 2003-07-05\\
1968L & N5236 &IIP&32.0& $< 1.5\times 10^{36}$ & ACIS & 2000-04-29, 2001-09-04\\ 
1969B & N3556 &IIP&32.6& $<3.8\times 10^{36}$& ACIS & 2001-09-08\\
1969L & N1058 &IIP&30.3& $<4.8\times 10^{37}$& ACIS & 2000-03-20\\
1970G & N5457 &IIL&33.9& $<4.9\times 10^{36}$& ACIS & 2004-07-05, 2004-07-11\\
1972Q & N4254 &IIP&30.5& $<3.0\times 10^{38}$& EPIC & 2003-06-29\\
1972R & N2841 &Ib &31.9& $< 7.3\times 10^{35}$& EPIC & 2004-11-09\\
1973R & N3627 &IIP&25.9& $<7.7\times 10^{37}$& ACIS & 1999-11-03\\
1976B & N4402 &Ib& 26.2& $<8.9\times 10^{37}$& EPIC & 2002-07-01\\
1979C & N4321 &IIL&26.8& $2.7^{+0.4}_{-0.4}\times 10^{38}$& ACIS & 2006-02-18\\
1980K & N6946 &IIL&24.0& $<6.5\times 10^{36}$& ACIS & 2004-10-22, 2004-11-06, 2004-12-03\\
1982F & N4490 &IIP&22.6& $<1.1\times 10^{36}$& ACIS & 2004-07-29, 2004-11-20\\
1983E & N3044 &IIL&19.0& $<4.6\times 10^{37}$& EPIC & 2001-11-24, 2002-05-10\\
1983I & N4051 &Ic &17.8 & $< 1.7\times 10^{36}$& ACIS & 2001-02-06\\
1983N & N5236 &Ib &16.8& $< 5.5\times 10^{36}$& ACIS & 2000-04-29\\
1983V & N1365 &Ic &19.1& $< 7.0\times 10^{37}$& ACIS & 2002-12-24\\
1985L & N5033 &IIL&14.9& $<8.1\times 10^{37}$& ACIS & 2000-04-28\\
1986E & N4302 &IIL&19.6& $1.4^{+0.5}_{-0.5}\times 10^{38}$& ACIS & 2005-12-05\\
1986I & N4254 &IIP&17.1& $<3.0\times 10^{38}$& EPIC & 2003-06-29\\
1986J & N891  &IIn&21.2 & $8.5^{+0.5}_{-0.5}\times 10^{38}$& ACIS & 2003-12-10\\
1986L & N1559 &IIL&18.9& $<1.4\times 10^{38}$& EPIC & 2005-08-10, 2005-10-12\\
1987B & N5850 &IIn&14.1& $< 1.5\times 10^{38}$& EPIC & 2001-01-25, 2001-08-26\\
1988A & N4579 &IIP&12.3& $<2.4\times 10^{37}$& ACIS & 2000-05-02\\
1988Z & MCG+03-28-22 &IIn&15.5& $2.9^{+0.5}_{-0.5}\times 10^{39}$& ACIS & 2004-06-29\\
1990U & N7479 & Ic & 10.9 & $1.1^{+0.6}_{-0.5}\times 10^{39}$& EPIC & 2001-06-19\\
1991N & N3310 &Ib/Ic &11.8& $<4.2\times 10^{37}$& ACIS & 2003-01-25\\
1993J & N3031 & IIb & 8.1 & $< 1.0 \times 10^{38}$& ACIS & 2001-04-22\\
1994I & N5194& Ic & 8.2 & $8.0^{+0.3}_{-0.7}\times 10^{36}$& ACIS & 2000-06-20, 2001-06-23, 2003-08-07\\
1994ak & N2782 &IIn&7.4& $< 3.7\times 10^{37}$& ACIS & 2002-05-17\\
1995N & MCG-02-38-17 &IIn&8.9& $4.3^{+1.0}_{-1.0}\times 10^{39}$& ACIS & 2004-03-27\\
1996ae & N5775 &IIn&5.9& $< 6.1\times 10^{37}$& ACIS & 2002-04-05\\
1996bu & N3631 &IIn&6.6& $<2.1\times 10^{37}$& ACIS & 2003-07-05\\
1996cr & ESO97-G13 &IIn&4.2& $1.9^{+0.4}_{-0.4}\times 10^{39}$& ACIS & 2000-03-14\\
1997X & N4691 &Ic&6.1& $< 2.2\times 10^{37}$& ACIS & 2003-03-08\\
1997bs & N3627 &IIn&2.5& $<2.9\times 10^{38}$& ACIS & 1999-11-03\\
1998S & N3877 &IIn&3.6& $3.8^{+0.5}_{-0.5}\times 10^{39}$& ACIS & 2001-10-17\\
1998T & N3690 &Ib&5.2& $< 2.0\times 10^{38}$& ACIS & 2003-04-30\\
1998bw & ESO184-G82 & Ic & 3.5 & $4.0^{+1.0}_{-0.9}\times 10^{38}$& ACIS & 2001-10-27\\
1999dn & N7714 &Ib&4.4& $<5.9\times 10^{37}$& ACIS & 2004-01-25\\
1999ec & N2207 &Ib& 5.9& $3.1^{+0.4}_{-0.4}\times 10^{39}$& EPIC & 2005-08-31\\
1999el & N6951 &IIn&5.6& $< 5.6\times 10^{38}$& EPIC & 2005-04-30, 2005-06-05\\
1999em & N1637 &IIP&1.0& $<1.4 \times 10^{37}$& ACIS & 2000-10-30\\
1999gi & N3184 &IIP&0.10& $2.6^{+0.6}_{-0.6}\times 10^{37}$& ACIS & 2000-01-08, 2000-02-03\\
2000P & N4965 &IIn&7.2& $< 1.2\times 10^{39}$& XRT & 2007-05-16\\
2000bg & N6240 &IIn&1.3& $<1.4\times 10^{39}$& ACIS & 2001-07-29\\
2001ci & N3079 &Ic &2.5& $< 5.0\times 10^{37}$& EPIC & 2003-10-14\\
2001du & N1365 &IIP&1.3& $<3.8\times 10^{37}$& ACIS & 2002-12-24\\
2001em & UGC11794 &Ib/Ic & 4.7 & $5.8^{+1.2}_{-1.2}\times 10^{40}$& EPIC & 2006-06-14\\
2001gd & N5033 & IIb&1.1 & $1.0^{+0.3}_{-0.3}\times 10^{39}$& EPIC & 2002-12-18\\
2001ig & N7424 &IIb&0.50 &$3.5^{+2.0}_{-2.0}\times 10^{37}$& ACIS & 2002-06-11\\

\end{tabular}
\end{table*}

\begin{table*}
\begin{tabular}{lllllll}
SN & Host galaxy & Type & Age (yr) & $L_{2-10\;keV} (\rm erg/s)$& Instrument & Observation date\\
\hline
2002ap & N628 &Ic&0.92& $< 3.1\times 10^{36}$& EPIC & 2003-01-07\\
2002fj & N2642 &IIn&4.7& $<1.3\times 10^{39}$& XRT & 2007-05-11\\
2002hf & MCG-05-3-20 &Ic&3.1& $<7.6\times 10^{38}$& EPIC & 2005-12-19\\
2003L & N3506 &Ic&0.08 &$7.7^{+1.5}_{-1.5}\times 10^{39}$& ACIS & 2003-02-10\\
2003ao & N2993 &IIP&0.016& $5.6^{+1.2}_{-1.2}\times 10^{38}$& ACIS & 2003-02-16\\
2003bg & MCG-05-10-15 &Ic/IIb&0.33 & $5.3^{+1.3}_{-0.8}\times 10^{38}$& ACIS & 2003-06-22\\
2003dh & Anon. &Ic &0.71& $< 5.0 \times 10^{40}$& EPIC & 2003-12-12\\
2003jd & MCG-01-59-21 &Ic &0.041& $< 3.0 \times 10^{38}$& ACIS & 2003-11-10\\
2003lw & Anon. &Ic&0.34& $7.0^{+3}_{-3}\times 10^{40}$& ACIS & 2004-04-18\\
2004C & N3683 &Ic &3.1& $1.0^{+0.3}_{-0.2}\times 10^{38}$& ACIS & 2007-01-31\\
2004dj & N2403 &IIP&0.34& $1.1^{+0.3}_{-0.3}\times 10^{37}$& ACIS & 2004-12-22\\
2004dk & N6118 &Ib& 0.030& $1.5^{+0.4}_{-0.4}\times 10^{39}$& EPIC & 2004-08-12\\
2004et & N6946 & IIP & 0.18 & $1.0^{+0.2}_{-0.2}\times 10^{38}$& ACIS & 2004-10-22, 2004-11-06, 2004-12-03\\  
2005N & N5420 &Ib/Ic &0.50 & $< 1.0\times 10^{40}$& XRT & 2005-07-17\\
2005U & Anon. &IIb &0.041& $< 1.1 \times 10^{39}$& ACIS & 2005-02-14\\
2005at & N6744 &Ic&1.7& $< 3.0\times 10^{38}$& XRT & 2006-10-31\\
2005bf & MCG+00-27-5 &Ib&0.58& $< 6.0\times 10^{39}$& XRT & 2005-11-07\\
2005bx & MCG+12-13-19 &IIn&0.25& $<1.0\times 10^{39}$& ACIS & 2005-07-30\\
2005da & UGC11301 &Ic&0.098& $<5.0\times 10^{39}$& XRT & 2005-08-23\\
2005db & N214 &IIn&0.036& $<2.0\times 10^{39}$&EPIC & 2005-08-01\\
2005ek & UGC2526 &Ic &0.041& $< 4.0\times 10^{39}$& XRT & 2005-10-07\\
2005gl & N266 &IIn&1.6& $<3.4\times 10^{39}$& XRT & 2007-06-01\\
2005kd & Anon. &IIn&1.2& $2.6^{+0.4}_{-0.4}\times 10^{41}$& ACIS & 2007-01-24\\
2006T & N3054 &IIb &0.0082& $< 6.0\times 10^{39}$& XRT & 2006-02-02\\
2006aj & Anon. &Ic & 0.43& $< 7.0 \times 10^{39}$& XRT & 2006-07-25\\
2006bp & N3953 &IIP&0.058& $1.0^{+0.2}_{-0.2}\times 10^{38}$& EPIC & 2006-04-30\\  
2006bv & UGC7848 &IIn&0.0082& $<1.2\times 10^{39}$& XRT & 2006-05-01\\
2006dn & UGC12188 &Ib&0.033& $< 2.5\times 10^{40}$& XRT & 2006-07-17\\
2006gy & N1260 &IIn&0.16& $<2.0\times 10^{38}$& ACIS & 2006-11-15\\
2006jc & UGC4904 &Ib&0.068& $2.1^{+0.6}_{-0.6}\times 10^{38}$& ACIS & 2006-11-04\\
2006lc & N7364 &Ib/Ic &0.016& $< 2.0\times 10^{40}$& XRT & 2006-10-27\\
2006lt & Anon. &Ib&0.068& $< 4.0\times 10^{39}$& XRT & 2006-11-05\\
2007C & N4981 &Ib &0.022& $<3.0\times 10^{40}$& XRT & 2007-01-15\\
2007D & UGC2653 &Ic &0.025& $< 3.0\times 10^{40}$& XRT & 2007-01-18\\
2007I & Anon. &Ic &0.016& $< 9.0\times 10^{39}$& XRT & 2007-01-20\\
2007bb & UGC3627 &IIn&0.022& $<4.4\times 10^{39}$& XRT & 2007-04-10\\

\hline

\end{tabular}
\caption{X-ray measurements and upper limits for our sample of
  historical supernovae. When more than one observation was used 
for a given source, the age is an average of three epochs weighted 
by their exposure lengths. The instrument ACIS is on-board {\em Chandra}, EPIC on {\em XMM-Newton}
and XRT on {\em Swift}.}
\end{table*}

We retrieved the relevant X-ray datasets from the public archives of
those three missions. The optical position of each SN in our sample is
well known, to better than $1\arcsec$: this makes it easier to
determine whether a SN is detected in the X-ray band (in particular
for {\it Chandra}), even with a very low number of counts, at a level
that would not be considered significant for source detection in a
blind search. For the {\it Chandra} observations, we applied standard
data analysis routines within the Chandra Interactive Analysis of
Observations ({\small CIAO}) software
package\footnote{http://cxc.harvard.edu/ciao} version 3.4.  Starting
from the level-2 event files, we defined a source region (radius
$2\farcs5$, comprising $\approx 95\%$ of the source counts at 2 keV,
on axis, and proportionally larger extraction radii for off-axis
sources) and suitable background regions not contaminated by other
sources and at similar distances from the host galaxy's nucleus. For
each SN, we extracted source and background counts in the $0.3$--$8$
keV band with {\it dmextract}.  In most cases, we are dealing with a
very small number of counts (e.g., 2 or 3, inside the source
extraction region) and there is no excess of counts at the position of
the SN with respect to the local background. In these cases, we
calculated the 90\% upper limit to the number of net counts with the
Bayesian method of Kraft et al. (1991). We then converted this net
count-rate upper limit to a flux upper limit with
WebPimms\footnote{http://heasarc.nasa.gov/Tools/w3pimms.html and
http://cxc.harvard.edu/toolkit/pimms.jsp  },
assuming a power-law spectral model with photon index $\Gamma = 2$ and
line-of-sight Galactic column density.  The choice of a power-law
spectral model is motivated by our search for X-ray emission from an
underlying pulsars rather than from the SN shock wave. In a few cases,
there is a small excess of counts at the SN position: we then also
built response and auxiliary response functions (applying {\it
psextract} in {\small CIAO}), and used them to estimate a flux,
assuming the same spectral model. When possible, for sources with
$\approx 20$--$100$ net counts, we determined the count rates
separately in the soft ($0.3$--$1$ keV), medium ($1$--$2$ keV) and
hard ($2$--$8$ keV) bands, and used the hard-band rates (essentially
uncontaminated by soft thermal-plasma emission, and unaffected by the
uncertainty in the column density and by the degradation of the ACIS-S
sensitivity) alone to obtain a more stringent value or upper limit to
the non-thermal power-law emission. Very few sources have enough
counts for a two-component spectral fit (mekal thermal plasma plus
power-law): in those cases, we used the 2-10 keV flux from the
power-law component alone in the best-fitting spectral model. For those
spectral fits, we used the {\small XSPEC} version 12 software package
(Arnaud 1996).

When we had to rely on {\it XMM-Newton}/EPIC data, we followed
essentially the same scheme: we estimated source and background count
rates (this time, using a source extraction circle with a $20\arcsec$
radius) in the full EPIC pn and MOS bands ($0.3$--$12$ keV) and, when
possible, directly in the $2$--$10$ keV band. The count rate to flux
conversion was obtained with WebPimms (with a $\Gamma=2$ power-law
model absorbed by line-of-sight column density) or through full
spectral analysis for sources with enough counts.  We used standard
data analysis tasks within the Science Analysis System ({\small SAS})
version 7.0.0 (for example, {\it xmmselect} for source extraction).
All three EPIC detectors were properly combined, both when we
estimated count rates, and when we did spectral fitting, to increase
the signal-to-noise ratio.  In fact, in almost all cases in which a
source position had been observed by both {\it Chandra} and {\it
XMM-Newton}, {\it Chandra} provided a stronger constraint to the flux,
because of its much narrower point-spread function and lower
background noise. The {\it Swift} data were analyzed using the Swift
Software version 2.3 tools and latest calibration products. Source
counts were extracted from a circular region with an aperture of
$20\arcsec$ radius centered at the optical positions of the SNe.  In
some cases, Swift observations referred to a Gamma ray burst (GRB)
associated to a core-collapse SN: we did not obviously consider the
GRB flux for our population analysis. Instead, for those cases, we
considered the most recent {\it Swift} observation after the GRB had
faded, and used that to determine an upper limit to a possible pulsar
emission. We only considered {\it Swift} observations deep enough to
detect or constrain the residual luminosity to $\la 10^{40}$ erg
s$^{-1}$.

In some cases, two or more {\it Chandra} or {\it XMM-Newton}
observations of the same SN target were found in the archive.  If they
were separated by a short interval in time (much shorter than the time
elapsed from the SN explosion), we merged them together, to increase
the detection threshold. The reason we can do this is that we do not
expect the underlying pulsar luminosity to change significantly
between those observations. However, when the time elapsed between
observations was comparable to the age of the SN, we attributed
greater weight to the later observations, for our flux
estimates.  The reason is that the thermal X-ray emission from the
shocked gas tends to decline more rapidly (over a few months or years)
than the non-thermal pulsar emission (timescale $\ga$ tens of
years). More details about the data analysis and the luminosity and
color/spectral properties of individual SNe in our sample will be
presented elsewhere (Pooley et al. 2008, in preparation). Here, we are
mainly interested in a population study to constrain the possible
presence and luminosity of high-energy pulsars detectable in the
$2$--$10$ keV band.

While this is, to the best of our knowledge, the first X-ray
search for pulsar wind nebulae (PWNe) in extragalactic SNe, and the first work that uses
these data to set statistical constraints on the properties of the
embedded pulsars, it should be noted that the possibility of
observing pulsars in young SNe (a few years old) was originally
discussed, from a theoretical point of view, by Chevalier \& Fransson
(1992).  Furthermore, searches for PWNe in extragalactic SNe have been
performed in the radio (Reynolds \& Fix 1987; Bartel \& Bietenholz
2005).  Observationally, however, clear evidence for pulsar activity
in SNe has been lacking.  The radio emission detected in some SNe,
although initially ascribed to pulsar activity (Bandiera, Pacini \&
Salvati 1984), was later shown to be well described as the result of
circumstellar interaction (Lundqvist \& Fransson 1988). There is
however a notable exception, that is SN 1986J, for which the observed
temporal decline of the H$_\alpha$ luminosity (Rupen et al. 1987) has
been considered suggestive of a pulsar energy input (Chevalier 1987).
As noted by Chevalier (1989), a possible reason for the apparent
low-energy input in some cases could be the fact that the embedded
neutron stars were born with a relatively long period. The present
work allows us to make a quantitative assessment on the typical
minimum periods allowed for the bulk of the NS population.
The list of SNe, their measured fluxes and their ages (at the time
of observation) are reported in Table~1. 

\section{Theoretical expectations for the X-ray luminosity of young pulsars}

Most isolated neutron stars are X-ray emitters throughout all their
life: at early times, their X-ray luminosity is powered by rotation
(e.g. Michel 1991; Becker \& Trumper 1997); after an age of $\sim 10^3
- 10^4$ yr, when the star has slowed down sufficiently, the main X-ray
source becomes the internal heat of the star\footnote{In the case
of magnetars, this internal heat is provided by magnetic field decay,
which dominates over all other energy losses.}, and finally, when this
is exhausted, the only possible source of X-ray luminosity would be
accretion by the interstellar medium, although to a very low
luminosity level, especially for the fastest stars (e.g. Blaes \&
Madau 1993; Popov et al. 2000; Perna et al. 2003).  Another possible
source of X-ray luminosity that has often been discussed in the
context of NSs is accretion from a fallback disk (Colgate 1971; Michel
\& Dressler 1981; Chevalier 1989; Yusifov et al. 1995; Chatterjee et
al. 2000; Alpar 2001; Perna et al. 2000).  Under these circumstances,
accretion would turn off magnetospheric emission, and X-ray radiation
would be produced as the result of accretion onto the surface of the
star. For a disk to be able to interfere with the magnetosphere and
accrete, the magnetospheric radius $R_m\sim 6.6\times 10^7
B_{12}^{4/7} \dot{m}^{-2/7}$~cm (with $\dot{m}^{-2/7}$ being the
accretion rate in Eddington units, and $B_{12}\equiv B/(10^{12} {\rm
G})$) must be smaller than the corotation radius $R_{\rm cor}\sim
1.5\times 10^8 P^{2/3} (M/M_\odot)^{1/3}$ cm.  If, on the other hand,
the magnetospheric radius resides outside of the corotation radius,
the propeller effect (Illarionov \& Sunyaev 1975) takes over and
inhibits the penetration of material inside the magnetosphere, and
accretion is (at least largely) suppressed.  For a typical pulsar
magnetic field $B_{12}\sim 5$, the magnetospheric radius becomes
comparable to the corotation radius for a period $P\sim 1$ s and an
Eddington accretion rate. If the infalling material does not possess
sufficient angular momentum, however, a disk will not form, but infall
of the bound material from the envelope is still likely to proceed,
albeit in a more spherical fashion. The accretion rate during the
early phase depends on the details of the (yet unclear) explosion
mechanism. Estimates by Chevalier (1989) yield values in the range
$3\times 10^{-4}$--$2\times 10^2 M_\odot$ yr$^{-1}$.  In order for the
pulsar mechanism to be able to operate, the pressure of the pulsar
magnetic field must overcome that of the spherical infall. For the
accretion rates expected at early times, however, the pressure of the
accreting material dominates over the pulsar pressure even at the
neutron star surface.  Chevalier (1989) estimates that, for accretion
rates $\dot{M}\ga 3\times 10^{-4}M_\odot$ yr$^{-1}$, the photon
luminosity is trapped by the inflow and the effects of a central
neutron star are hidden.  Once the accretion rate drops below that
value, photons begin to diffuse out from the shocked envelope; from
that point on, the accretion rate drops rapidly, and the pulsar
mechanism can turn on. Chevalier (1989) estimates that this occurs at
an age of about 7 months.  Therefore, even if fallback plays a major
role in the initial phase of the SN and NS lives, its effects are not
expected to be relevant at the timescales of interest for the
conclusions of this work.

For the purpose of our analysis, we are especially interested in the
X-ray luminosity at times long enough so that accretion is
unimportant, but short enough that rotation is still the main source
of energy.  During a Crab-like phase, relativistic particles
accelerated in the pulsar magnetosphere are fed to a synchrotron
emitting nebula, the emission of which is characterized by a powerlaw
spectrum. Another important contribution is the pulsed X-ray
luminosity (about 10\% of the total in the case of the Crab)
originating directly from the pulsar magnetosphere. It should be noted
that one important assumption of our analysis is that all (or at least
the greatest majority) of neutron stars goes through an early time
phase during which their magnetosphere is active and converts a
fraction of the rotational energy into X-rays.  However, there is
observational evidence that there are objects, known as Central
Compact Objects\footnote{Examples are central source in Cas A and in
Puppis A (e.g. Petre et al. 1996; Pavlov et al. 2000).} (CCOs), for
which no pulsar wind nebulae are detected.  Since no pulsations are
detected for these stars, it is possible that they are simply objects
born slowly rotating and which hence have a low value of $\dot{E}_{\rm
rot}$.  In this case, they would not affect any of our considerations,
since the $L_x-\dot{E}_{\rm rot}$ correlation appears to hold all the
way down to the lowest measured values of $L_x$ and $\dot{E}_{\rm
rot}$.  However, if the CCOs are NSs with a high $\dot{E}_{\rm rot}$,
but for which there exists some new physical mechanism that suppresses
the magnetospheric activity (and hence the X-ray luminosity) to values
much below what allowed by the scatter in the pulsar $L_x-\dot{E}_{\rm
rot}$ relation, then these stars would affect the limits that we
derive. Since at this stage their nature is uncertain, we treat the
all sample of NSs on the same footing, although keeping this in mind
as a caveat should future work demonstrate the different intrinsic
nature of the CCOs with respect to the conversion of rotational energy
into X-ray luminosity.

As discussed in \S1, for all the neutron stars for which both the
rotational energy loss, $\dot{E}_{\rm rot}$, and the X-ray luminosity,
$L_x$, have been measured, there appears to be a correlation between
these two quantities.  This correlation appears to hold over a wide
range of rotational energy losses, including different emission
mechanisms of the pulsar. Since in the high $L_x$ regime (young
pulsars) of interest here the X-ray luminosity is dominated by
rotational energy losses, the most appropriate energy band for our
study is above $\sim 2$ keV, where the contribution of surface
emission due to the internal energy of the star is small. The
correlation between $L_x$ and $\dot{E}_{\rm rot}$ in the 2-10 keV band
was first examined by Saito et al. (1997) for a small sample of
pulsars, and a more comprehensive investigation with the largest
sample up to date was later performed by P02.  They found, for a
sample of 39 pulsars, that the best fit is described by the relation
\beq 
\log L_{x,[2-10]}= 1.34\,\log \dot{E}_{\rm rot} -15.34\;,
\label{eq:Lx}
\eeq with $1\sigma$ uncertainty intervals on the parameters $a=1.34$
and $b=15.34$ given by $\sigma_a=0.03$ and $\sigma_b=1.11$,
respectively.  A similar analysis on a subsample of 30 pulsars with
ages $\tau < 10^6$ yr by Guseinov et al. (2004) yielded a best fit with
parameters $a=1.56$, $b=23.4$, and corresponding uncertainties
$\sigma_a=0.12$ and $\sigma_b=4.44$. The slope of this latter fit is a
bit steeper than that of P02; as a result, the model by Guseinov et
al. predicts a larger fraction of high luminosity pulsars from the
population of fast rotating young stars with respect to the best fit
of P02.  In order to be on the conservative side for the predicted
number of high-$L_x$ pulsars, we will use as our working model the one
by P02.  It is interesting to note, however, that both groups
find that the efficiency $\eta_x\equiv L_x/\dot{E}_{\rm rot}$ is an
increasing function of the rotational energy loss $\dot{E}_{\rm rot}$
of the star. Furthermore, the analysis by Guseinov et al. shows that,
for a given $\dot{E}_{\rm rot}$, pulsars with larger $B$ field have a
systematically larger efficiency $\eta_x$ of conversion of rotational
energy into X-rays.  An increase of $\eta_x$ with $\dot{E}_{\rm
rot}$ was found also in the investigation by Cheng, Taam \& Wang
(2004). They considered a sample of 23 pulsars and studied the trend
with $\dot{E}_{\rm rot}$ of the pulsed and non-pulsed components
separately. Their best-fit yielded $L_x^{\rm pul}\propto\dot{E}_{\rm
rot}^{1.2\pm 0.08}$ for the pulsed component, and $L_x^{\rm
npul}\propto\dot{E}_{\rm rot}^{1.4\pm 0.1}$ for the non-pulsed
one. They noticed how the former is consistent with the theoretical
X-ray magnetospheric emission model by Cheng \& Zhang (1999), while
the latter is consistent with a PWN model in which $ L_x^{\rm
npul}\propto\dot{E}_{\rm rot}^{p/2}$, where $p\sim 2-3$ is the
powerlaw index of the electron energy distribution. Their best fit for
the total X-ray luminosity (pulsed plus unpulsed components) yielded
$L_x\propto\dot{E}_{\rm rot}^{1.35\pm 0.2}$, fully consistent with
the best-fit slope of P02.  Along similar lines, recently Li et
al. (2007) presented another statistical study in which, using {\em
Chandra} and {\em XMM} data of galactic sources, they were able to
resolve the component of the X-ray luminosity due to the pulsar from
that due to the PWN. Their results were very
similar to those of Cheng et al. (2004), with a best fit for the
pulsar component $L_x^{\rm psr}\propto\dot{E}_{\rm rot}^{1\pm 0.1}$,
and a best fit for the PWN (representing the unpulsed contribution)
$L_x^{\rm PWN}\propto\dot{E}_{\rm rot}^{1.4\pm 0.2}$. They found that
the main contribution to the total luminosity generally comes from the
unpulsed PWN, hence yielding the steepening of the $L_X-\dot{E}_{\rm
rot}$ correlation with $\dot{E}_{\rm rot}$, consistently with the P02
relation, where the contribution from the pulsar and the PWN are not
distinguished. For our purposes, we consider the sum of both
contributions, since we cannot resolve the two components in the
observed sample of historical SNe.\footnote{For typical distances $\ga$
a few Mpc, the ACIS spatial resolution is $\ga 20$ pc, while 
PWN sizes are on the order of a fraction of a pc to a few pc.}

It should be noted that, despite the general agreement among the
various studies on the trend of $L_x$ with $\dot{E}_{\rm rot}$, and
the support from theory that the correlation is expected to steepen
with $\dot{E}_{\rm rot}$, there must be a point of saturation in order
to always satisfy the condition $L_x \le \dot{E}_{\rm rot}$. While in
our simulations we impose the extra condition that $\eta_x\le 1$, it
is clear that, until the correlation can be calibrated through direct
measurements of objects with high values of $\dot{E}_{\rm rot}$, there
remains an uncertainty on how precisely the saturation occurs, and
this uncertainty is unavoidably reflected in the precise details of
our predictions.  However, unless there is, for some reason, a
point of turnover above the observed range where the efficiency of
conversion of rotational energy into X-rays turns back into $\eta_x\ll
1$, then our general conclusions can be considered robust.
In our analysis, in order to quantify the uncertainty associated
with the above, we will also explore the consequences of a break in
$\eta_x$ just above the observed range (to be mostly conservative).

Another point to note is that one implicit assumption that we make in
applying the $L_x-\dot{E}_{\rm rot}$ relation to very young objects is
that the synchrotron cooling time $t_{\rm synch}$ in X-rays is much
smaller than the age of the source, so that the X-ray luminosity is
essentially an instantaneous tracer of $\dot{E}_{\rm rot}$. In order
to check the validity of this assumption, we have made some rough
estimates based on measurements in known sources.  For example, let's
consider the case of the PWN in SN 1986J. Although the field in the
PWN has not been directly measured, we can use radio equipartition and
scale it from that of the Crab Nebula.  The Crab's radio synchrotron
emission has a minimum energy of $\sim 6\times 10^{48}$ ergs (see
e.g. Chevalier 2005), and a volume of $\sim 5\times 10^{56}$
cm$^3$. The average magnetic field is then $\sim 550$ $\mu$G.  We can
then scale to the PWN in SN 1986J, using the fact that the radio
luminosity is related to that of the Crab by $L_{\rm r, 1986J} \sim
200 L_{\rm r,Crab}$, while its size is about 0.01 times that of the
Crab. According to equipartition, $B_{\rm min} \sim ({\rm
size})^{-6/7} L_r^{2/7}$ (e.g. Willott et al. 1999), so this very
crude approach suggests that the magnetic field in the PWN of SN 1986J
is $B_{\rm min} \sim 235 B_{\rm Crab} \sim 120$ mG. This yields a very
short cooling time in X-rays, $t_{\rm synch}\sim 5$ hr (assuming a
Lorentz factor for the electrons of $\sim 10^6$), so that, if we scale
from the Crab nebula, the use of $L_x-\dot{E}_{\rm rot}$ at early
times appears reasonable.  If, on the other hand, initial periods are
generally slower than for the Crab, then the equipartition energy
could be much smaller and the corresponding lifetimes much
longer. Let's then consider a $10^{12}$ G pulsar with an initial
period of 60 ms (the pulsar produced in SN 386 is such a source). We
then have $\dot{E}_{\rm rot} \sim 3\times 10^{36}$ erg/s, so that
(ignoring expansion losses), the energy deposited in the PWN over 20
years would be $\sim 2\times 10^{45}$ ergs. For a volume similar to
that for SN 1986J above, the equipartition magnetic field would be
$\sim10$ mG, corresponding to a lifetime at 2 keV of about 10
days. This is still a short enough lifetime for our purposes.
Alternatively, for $P_0\sim 5$ ms and $B \sim 10^{12}$ G, we have
$\dot{E}_{\rm rot} \sim 6\times 10^{40}$ erg/s, and over 20 years,
this yields $E_{\rm tot} \sim 4\times 10^{49}$ ergs.  This implies $B
> 1$~ G, so that $t_{\rm synch} \sim 15$ minutes.  Therefore, we
conclude that, overall, the magnetic fields in young PWNe are likely
strong enough to justify the use of the $L_x-\dot{E}_{\rm rot}$
relation even for the youngest objects in our sample.

One further point to notice with respect to the $L_x-\dot{E}_{\rm
rot}$ correlation is the fact that it is based on a diversity of
objects. The low end range of the relation, in particular, is
populated with Millisecond Pulsars (MSPs), which are spun up neutron
stars. It is possible that this class of objects might bias the
correlation of the youngest, isolated pulsars in the sample.
Generally speaking, once they are spun up, the MSPs form PWNe again
(e.g. Cheng et al. 2006), and the conversion of $\dot{E}_{\rm rot}$
into $L_x$, which is practically an instantaneous relationship (as
compared to the ages under consideration), should not be dependent on
the history of the system. The magnetic field of the objects (lower
for the MSPs than for the young, isolated pulsars), however, might
influence the conversion efficiency (Guseinov et al. 2004), hence
biasing the overall slope of the correlation. Overall, in our analysis
a steeper slope would lead to tighter limits on the NS spin birth
distribution, and viceversa for a shallower slope. What would be
affected the most by a slope change is the high $\dot{E}_{\rm rot}$
tail of the population.  Hence, in \S4, besides deriving results using
the $L_x-\dot{E}_{\rm rot}$ for all the pulsars, we will also examine
the effects of a change of slope for the fastest pulsars.

The rotational energy loss of the star, under the assumption that it is dominated
by magnetic dipole losses, is given by
\begin{equation}
\dot{E}_{\rm rot}= \frac{B^2\sin^2\theta\,\Omega^4\,R^6}{6c^3}\;,
\label{eq:Edot}
\end{equation}

where $R$ is the NS radius, which we take to be 10 km, 
$B$ the NS magnetic field, $\Omega=2\pi/P$ the star angular velocity, and 
$\theta$ the angle between the magnetic and spin axes. We take $\sin\theta=1$
for consistency with what generally assumed in pulsar radio studies.
  With $\sin\theta=1$ and a
constant $B$ field, the spin evolution of the pulsars is simply given
by
\beq P(t) = \left[P_0^2 + \left(\frac{16\pi^2 R^6
B^2}{3Ic^3}\right)t\right]^{1/2}\;,
\label{eq:spin} 
\eeq 
where $I\approx 10^{45} {\rm g}\, {\rm cm}^2$ is the moment of
inertia of the star, and $P_0$ is its initial spin period.  The
X-ray luminosity of the pulsar at time $t$ (which traces $\dot{E}_{\rm
rot}$) correspondingly declines as $ L_x=L_{x,0}(1+t/t_0)^{-2}$, where
$t_0\equiv 3Ic^3P_0^2/B^2R^6(2\pi)^2 \sim 6500\, {\rm
yr}\;I_{45}B^{-2}_{12}R_{10}^{-6}P^2_{0,10}$, having defined
$I_{45}\equiv I/(10^{45}$~g~cm$^2$), $R_{10}\equiv R/(10\; {\rm km})$,
$P_{0,10}\equiv P_0/(10\;{\rm ms})$.  For $t\la t_0$ the flux does not
vary significantly.  Since the ages $t_{\rm SN}$ of the SNe in our
sample are all $\la 77$ yr, we deduce that, for typical pulsar fields, $t_{\rm SN}\ll
t_0$. The luminosities of the
pulsars associated with the SNe in our sample are therefore expected to be
still in the plateau region, and thus they directly probe the initial birth
parameters, before evolution affects the periods appreciably.

In order to compute the X-ray luminosity distribution of a population
of young pulsars, the magnetic fields and the initial periods of the
pulsars need to be known.  As discussed in \S1, a number of
investigations have been made over the last few decades in order to
infer the birth parameters of NSs, and in particular the distribution
of initial periods and magnetic fields. Here, we begin our study by
comparing the SN data with the results of a pulsar population
calculation that assumes one of such distributions, and 
specifically one that makes predictions for birth periods in the
millisecond range.  After establishing that the SN data are highly
inconsistent with such short initial spins, we then generalize our
analysis by inverting the problem, and performing a parametric study
aimed at finding the minimum values of the birth periods that result
in predicted X-ray luminosities consistent with the SN X-ray data.

\section{Observational constraints on the pulsar X-ray luminosities from 
comparison with historical SNe}

As a starting point to constrain pulsar birth parameters, we consider
the results of one of the most recent and comprehensive radio studies,
based on large-scale radio pulsar surveys, that is the one carried out
by Arzoumanian et al. (2002; ACC in the following). They find that, if
spin down is dominated by dipole radiation losses (i.e. braking index
equal to 3), and the magnetic field does not appreciably decay, the
magnetic field strength (taken as Gaussian in log) has a mean $\langle
\log B_0[G] \rangle =12.35$ and a standard deviation of $0.4$, while
the initial birth period distribution (also taken as a log-Gaussian),
is found to have a mean $\langle \log P_0(s) \rangle =-2.3$ with a
standard deviation $\sigma_{P_0}> 0.2$ (within the searched range of
$0.1-0.7$).  In the first part of our paper, as a specific example of
a distribution that predicts a large fraction of pulsars to be born
with millisecond periods, we use their inferred parameters described
above.  Since in their model the standard deviation for the initial
period distribution is constrained only by the lower limit
$\sigma_{P_0}>0.2$, here we adopt $\sigma_{P_0}=0.3$. As the width of
the velocity dispersion increases, the predicted X-ray luminosity
distribution becomes more and more heavily weighed towards higher
luminosities (see Figure 1 in Perna \& Stella 2004). Therefore, the
limits that we derive in this section would be even stronger if
$\sigma_{P_0}$ were larger than what we assume.  We then assume that
the $L_X-\dot{E}_{\rm rot}$ correlation is described by the P02 best
fit with the corresponding scatter.

In order to test the resulting theoretical predictions for the pulsar
distribution of X-ray luminosities against the limits of the SNe, we
perform $10^6$ Monte Carlo realizations of the compact object remnant
population. Each realization is made up of $N_{\rm obj}=N_{\rm
SN}=100$, with ages equal to the ages of the SNe in our sample. The
fraction of massive stars that leave behind a black hole (BH) has been
theoretically estimated in the simulations by Heger et al. (2003).
For a solar metallicity and a Salpeter IMF, they find that this
fraction is about 13\% of the total.  However we need to remark that
while, following their predictions, in our Monte Carlo
simulations we assign a 0.13 probability for a remnant to contain a
BH, the precise BH fraction is, in reality, subject to a certain
degree of uncertainty.  Even taking rigorously the results of Heger et
al. (2003), one needs to note that their NS vs BH fraction (cfr. their
fig.5) was computed assuming a fraction of about 17\% of Type Ib/c
SNe, and 87\% of Type II. Our sample, on the other hand, contains
about 40\% of Type Ib/c and 60\% of Type II SNe. How the remnant
fraction would change in this case is difficult to predict.  Heger et
al. point out how normal Type Ib/c SNe are not produced by single
stars until the metallicity is well above solar. In this case, the
remnants would be all NSs. At lower metallicities, on the other hand,
most Type Ib/c SNe are produced in binary systems where the binary
companion helps in removing the hydrogen envelope of the collapsing
star.  Given these uncertainties, while adopting for our simulations
the BH/NS fraction estimated by Heger et al. for solar metallicity, we
also discuss how results would vary for different values of the BH and
NS components.

If an object is a BH, a low level of X-ray luminosity ($<10^{35}$
erg/s, i.e. smaller than the lowest measurement/limit in our SN data
set) is assigned to it. This is the most conservative assumption that
we can make in order to derive constraints on the luminosity
distribution of the NS component.  If an object is a NS, then its
birth period and magnetic field is drawn from the ACC distribution as
described above, and it is evolved to its current age (equal to the
age of the corresponding SN at the time of the observation) with
Eq.(\ref{eq:spin}).  The corresponding X-ray luminosity is then drawn
from a log-Gaussian distribution with mean given by the P02 relation,
and dispersion $\sigma_{L_x}=\sqrt{\sigma_a^2[\log\dot{E}_{\rm rot}]^2
+ \sigma_b^2}$.

Figure 1 (top panel) shows the predicted distribution of the most
frequent value\footnote{For each (binned) value of the pulsar
luminosity, we determined the corresponding probability distribution
resulting from the Monte Carlo simulations. The maximum of that
distribution is what we indicated as the ``most frequent'' value for
each bin.} of the pulsar luminosity over all the Monte Carlo
realizations of the entire sample of Table~1. The shaded region
indicates the $1\sigma$ dispersion in the model. This has been
determined by computing the most compact region containing 68\% of the
random realizations of the sample.  Also shown is the distribution of
the X-ray luminosity (both detections and upper limits) of the SNe
(cfr. Table~1).  Since the measured X-ray luminosity of each object is
the sum of that of the SN itself and that of the putative pulsar
embedded in it, for the purpose of this work X-ray detections are also
treated as upper limits on the pulsar luminosities.  This is indicated
by the arrows in Figure~1.

\begin{figure}
\psfig{file=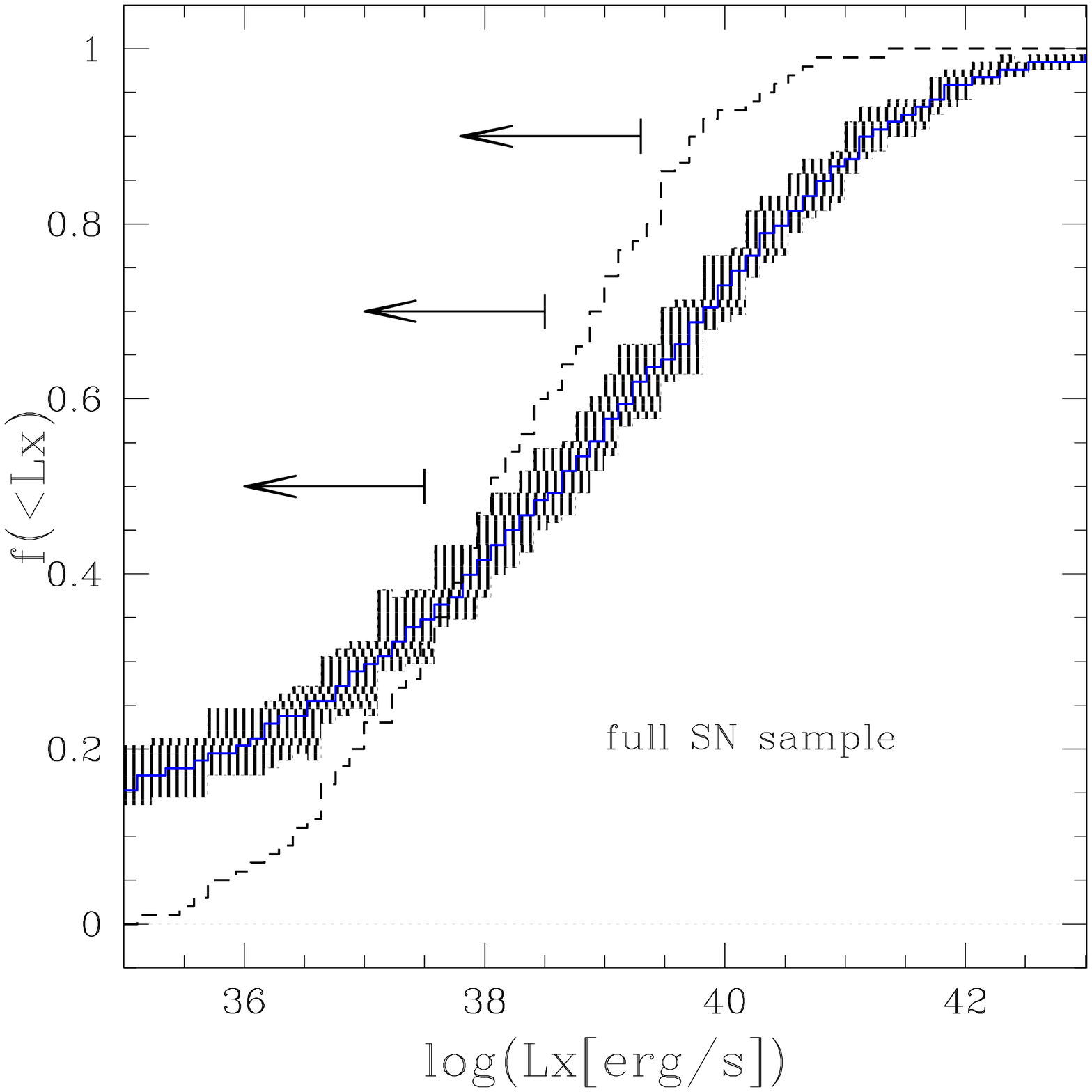,width=0.38\textwidth}
\vspace{-0.05in}
\psfig{file=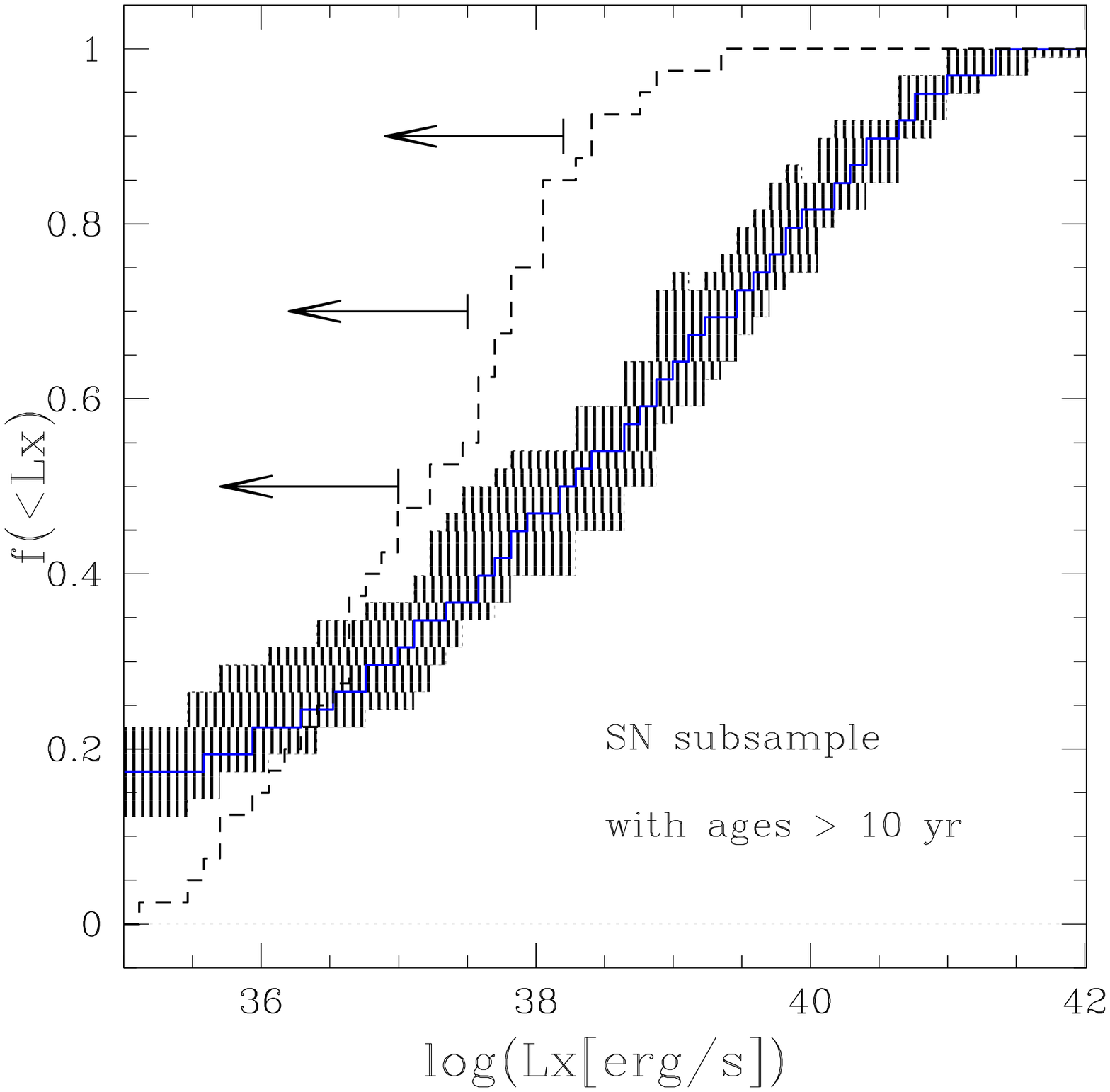,width=0.38\textwidth}
\vspace{-0.05in}
\psfig{file=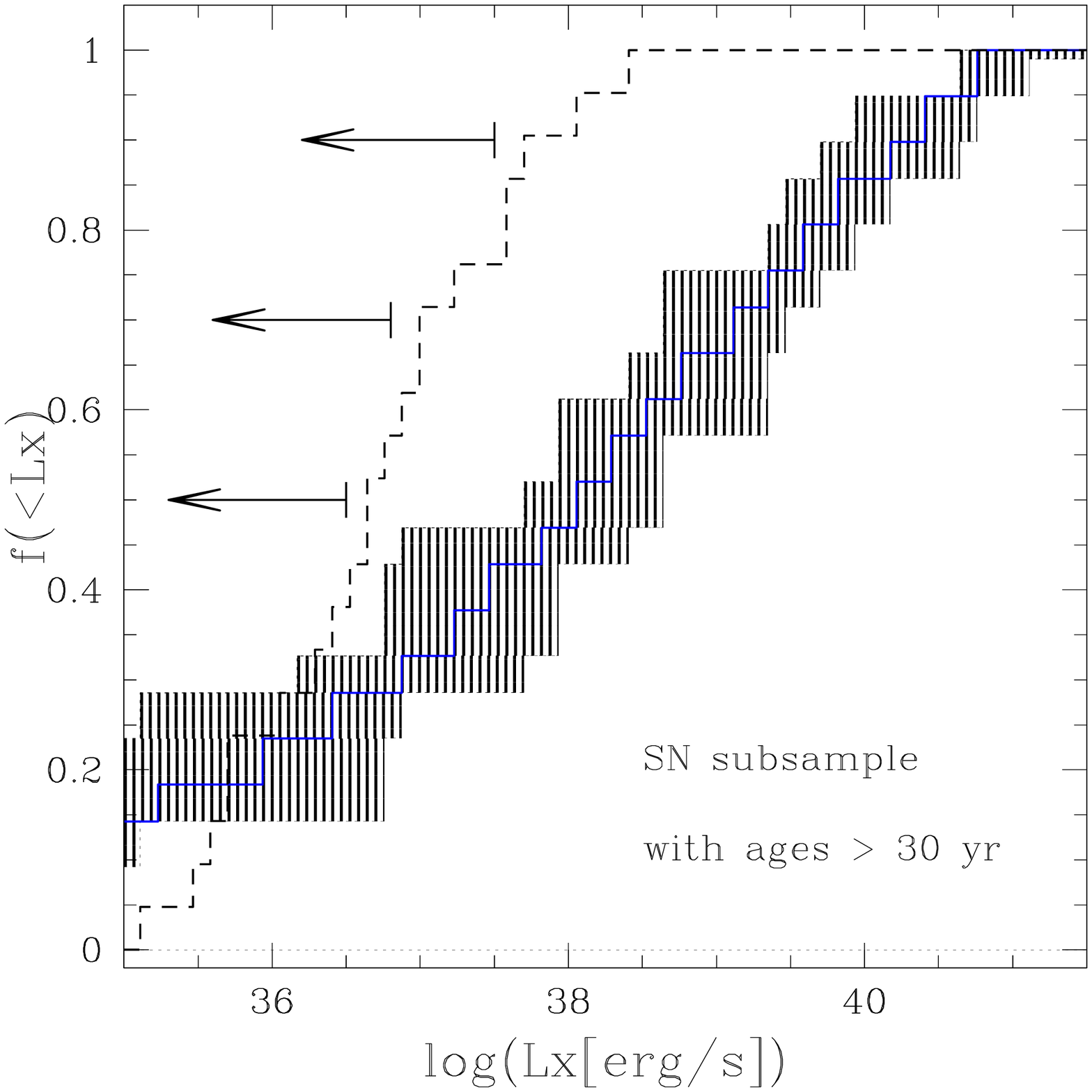,width=0.38\textwidth}
\vspace{-0.07in}
\caption{The dashed line shows the distribution of 2-10 keV luminosities
(either measurements or upper limits) for the entire sample of 100 SNe
analyzed {\em (upper panel)}, for the subsample of SNe with ages $>10$
yr {\em (middle panel)}, and with ages $>30$ yr {\em (lower panel)}.
The measured SN luminosities are also treated as upper limits on the
luminosities of the embedded pulsars.  The solid line shows the
prediction for the X-ray luminosity distribution of pulsars born
inside those SNe, according to the ACC birth parameters and the
$L_X-\dot{E}_{\rm rot}$ P02 relation.  The shaded regions indicate the
$1$-$\sigma$ confidence level of the model, derived from $10^6$ random
realizations of the sample. Independently of the SN sample considered,
the pulsar luminosity distribution is highly inconsistent with the
corresponding SN X-ray limits.}
\end{figure}

Our X-ray analysis, in all those cases where a measurement was
possible, never revealed column densities high enough to affect the
observed 2-10 keV flux significantly.  However, if a large fraction of
the X-ray luminosity (when not due to the pulsar) does not come from
the innermost region of the remnant, then the inferred $N_{\rm H}$
would be underestimated with respect to the total column density to
the pulsar.  The total optical depth to the center of the SN as
a function of the SN age depends on a number of parameters, the most
important of which are the ejected mass and its radial
distribution. The density profile of the gas in the newly born SN is
determined by the initial stellar structure, as modified by the
explosion. Numerical simulations of supernova explosions produce
density distributions that, during the free expansion phase, can be
approximated by the functional form $\rho_{\rm SN}=f(v)t^{-3}$ (see
e.g.  Chevalier \& Fransson 1994 and references therein). The function
$f(v)$ can in turn be represented by a power-law in velocity,
$f(v)\propto v^{-n}$.  To date, the best studied case is that of SN
1987A. Modeling by Arnett (1988) and Shigeyama \& Nomoto (1990)
yield an almost flat inner powerlaw region, surrounded by a very
steep outer powerlaw profile, $n\sim 9-10$. For normal abundances and
at energies below\footnote{Above 10 keV, the opacity is dominated by
electron scattering, which is energy independent.} 10 keV, Chevalier
\& Fransson (1994) estimate the optical depth at energy $E_{10}\equiv
E/(10\;{\rm keV})$ to the center of a supernova with a flat inner
density profile to be $\tau=\tau_s E_{10}^{-8/3}\; E_{\rm SN,
51}^{-3/2}\; M_{\rm ej, 10}^{5/2} t_{\rm yr}^{-2}$, where $E_{\rm SN,
51}$ is the supernova energy in units of $10^{51}$ erg, and $M_{\rm
ej,10}$ is the mass of the ejecta in units of 10 $M_\odot$.  The
constant $\tau_s$ is found to be 5.2 for a density profile with $n=7$
in the outer parts, and 4.7 for $n=12$. From these simple estimates,
it can be seen that the SN would have to wait a decade or so before
starting to become optically thin at the energies of interest.  These
estimates however do not account for the fact that, if the SN harbors
an energetic pulsar in its center, the pulsar itself will ionize a
substantial fraction of the surrounding neutral material. Calculations
of the ionization front of a pulsar in the interior of a young SN were
performed by Chevalier \& Fransson (1992). In the case of a flat
density profile in the inner region, and an outer density profile with
powerlaw $n=9$, they estimate that the ionization front reaches the
edge of the constant density region after a time $t_{\rm yr}=10\;t_0
f_i^{-1/3} \dot{E}_{\rm rot, 41}^{-1/3}\;M_{\rm ej, 10}^{7/6}\;E_{\rm
SN, 51}^{-1/2}$, where $\dot{E}_{\rm rot, 41}\equiv\dot{E}_{\rm
rot}/10^{41}\;{\rm erg}\;{\rm s^{-1}}$, and $f_i$ is the fraction of
the total rotational power that is converted in the form of ionizing
radiation with a mean free path that is small compared to the
supernova size. The constant $t_0$ depends of the composition of the
core. For a hydrogen-dominated core, $t_0=1.64$, for a
helium-dominated core, $t_0=0.69$, and for an oxygen-dominated core
$t_0=0.28$. Once the ionization front has reached the edge of the
constant density region, the steep outer power-law part of the
density profile is rapidly ionized.  Therefore, depending on the
composition and total mass of the ejecta, an energetic pulsar can
ionize the entire mass of the ejecta on a timescale between a few
years and a few tens of years. This would clearly reduce the optical
depth to the center of the remnant estimated above.

Given these considerations, in order to make predictions that are
not as likely to be affected by opacity effects, we also performed a
Monte Carlo simulation of the compact remnant population for all the
SNe with ages $t>10$ yr, and another for all the SNe with ages $t>30$~yr.
Since the opacity scales as $t^{-2}$, these subsets of objects
are expected to be substantially less affected by high optical depths
to their inner regions. The subsample of SNe with ages $t>10$~yr
contains 40 objects, while the subsample with ages $t>30$~yr contains
21 SNe. The corresponding luminosity distributions (both measurements
and limits) are shown in Figure~1 (middle and bottom panel
respectively), together with the predictions of the adopted model (ACC
initial period distribution and P02 $L_x-\dot{E}_{\rm rot}$
correlation) for the luminosities of the pulsars associated with those
SN samples. Given the uncertainties in the early-time optical depth,
we consider the constraints derived from these subsamples (and
especially the one with $t>30$ yr) more reliable.  Furthermore, even
independently of optical depth effects that can bias the youngest
members of the total sample, the subsamples of older SNe have on average
lower luminosities, hence making the constraints on the model
predictions more stringent. In the following, when generalizing our
study to derive limits on the allowed initial period distribution, we
will use for our analysis only the subsets of older SNe.

In all three panels of Figure~1, the low luminosity tail of the
simulation, accounting for $\sim 15\%$ of the population, is dominated
by the fraction of SNe whose compact remnants are black holes, and for
which we have assumed a luminosity lower than the lowest SN
measurement/limit ($\sim 10^{35}$ erg/s).  While it is possible that
newly born BHs could be accreting from a fallback disk and hence have
luminosities as high as a few $\times 10^{38}$ erg/s, our assumption
of low luminosity for them is the most conservative one for the
analysis that we are performing, in that it allows us to derive the
most stringent limits on the luminosity of the remaining remnant
population of neutron stars.  For these, the high-luminosity tail is
dominated by the fastest pulsars, those born with periods of a few
ms. The magnetic fields, on the other hand, are in the bulk range of
$10^{12}-10^{13}$ G.  The low-$B$ field tail produces lower
luminosities at birth, while the high-$B$ field tail will cause the
pulsars to slow down on a timescale smaller than the typical ages of
the SNe in the sample.  Therefore, it is essentially the initial
periods which play a crucial role in determining the extent of the
high-luminosity tail of the distribution.  With the birth parameter
distribution used here, we find that, out of the $10^6$ Monte Carlo
realizations of the sample (for each of the three cases of Fig.1),
none of them  predicts pulsar luminosities compatible with the SN
X-ray limits.\footnote{We need to point out that, in the
study presented here, we refrain from performing detailed probability
analysis. This is because, given the observational uncertainties of
some of the input elements needed for our study (as discussed  both
above and in the following), precise probability numbers would not be
especially meaningful at this stage.}

These results point in the direction of initial periods of the pulsar
population to be slower than the ms periods derived from some
population synthesis studies in the radio. A number of other
investigations in the last few years, based on different methods of
analysis of the radio sample with respect to ACC, have indeed come up
to conclusions similar to ours.  The population synthesis studies of
Faucher-Giguere \& Kaspi (2005) yielded a good fit to the data with
the birth period described by a Gaussian with a mean period of 0.3 s
and a spread of 0.15 sec.  Similarly, the analysis by Ferrario \&
Wickramasinghe (2006) yielded a mean period of 0.23 sec for a magnetic
field of $10^{12}$ G.  We performed Monte Carlo simulations of the
X-ray pulsar population using the birth parameters derived in those
studies above, and found them to be consistent with the SNe X-ray
limits shown in Figure~1.

In order to generalize our analysis beyond the testing of known
distributions, we performed a number of Monte Carlo simulations with
different initial spin period distributions and a mean magnetic field
given by the optimal model of Faucher-Giguere \& Kaspi (2006).  This
is a log-Gaussian with mean $\left<\log(B/{\rm G})\right>=12.65$ and
dispersion $\sigma_{\log B}=0.55$.\footnote{The inferred values of the
magnetic field in different studies are all generally in this range,
even for very different inferred spin birth parameters. Furthermore,
Ferrario \& Wickramasinghe (2006) note that the pulsar birth period
that they infer is almost independent of the field value in the range
$\log B({\rm G})=10-13$, where the vast majority of isolated radio
pulsar lie.}.

\begin{figure}
\psfig{file=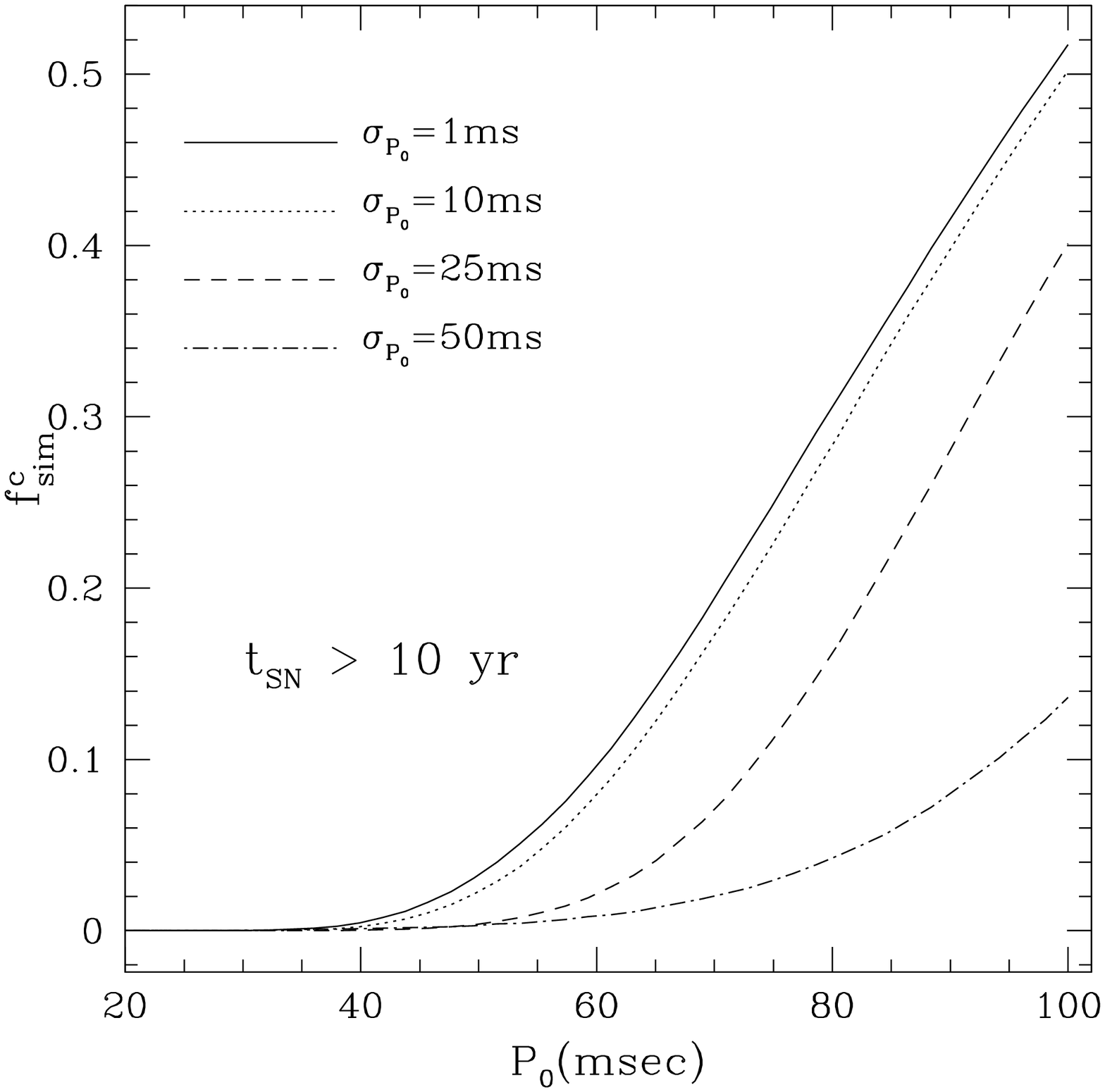,width=0.48\textwidth}
\psfig{file=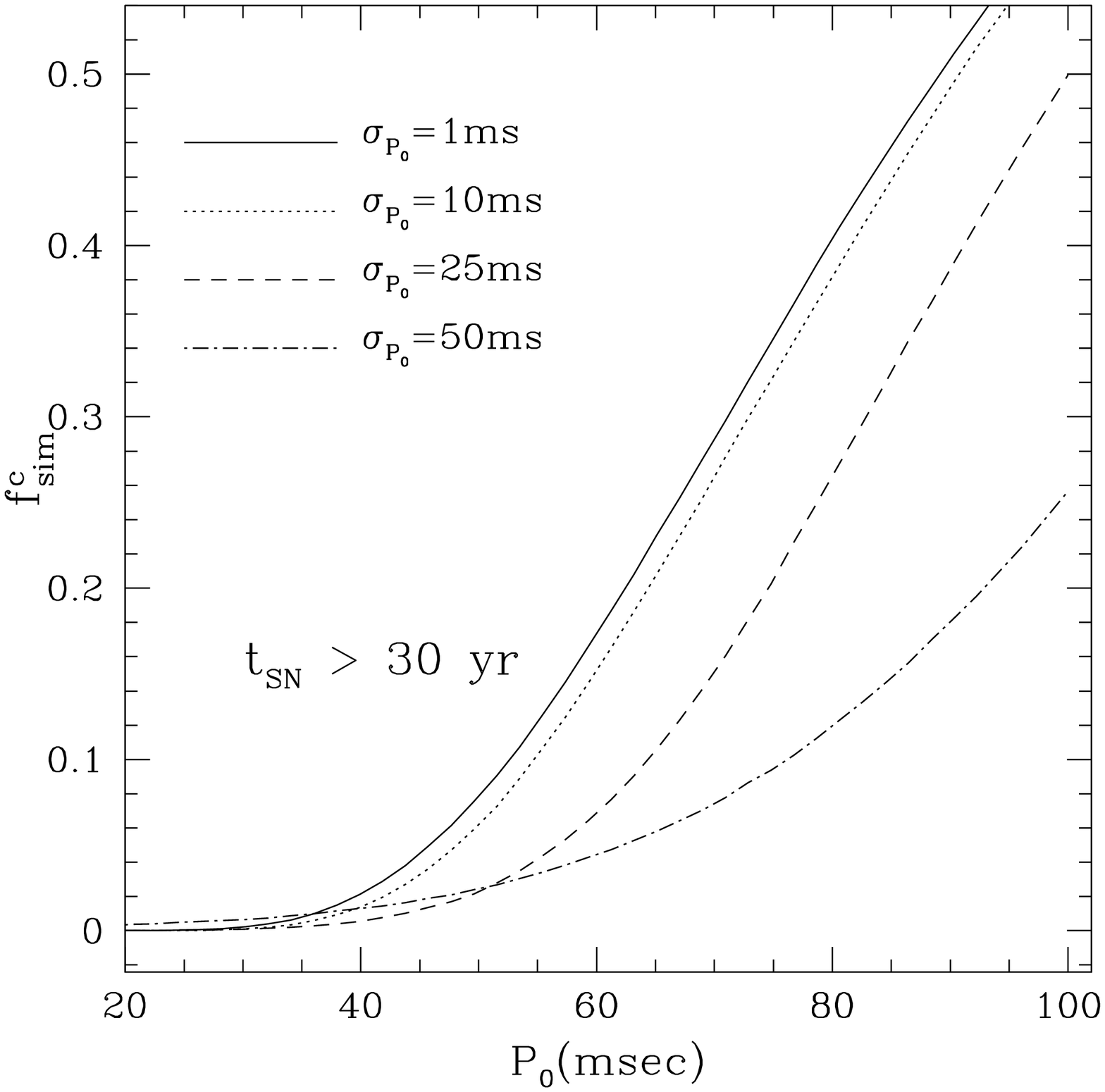,width=0.48\textwidth}
\caption{Fraction $f_{sim}^c$ of Monte Carlo realizations of the SN
sample for which the 2-10 keV luminosities of the pulsars are below
the limits of the corresponding SNe. This is shown for different
distributions of the initial spin periods, described by Gaussians of
mean $P_0$ and dispersion $\sigma_{P_0}$. In the upper panel, the
sample includes only the SNe of ages $>10$ yr (cfr. Fig.1, middle
panel), while in the lower panel only the SNe with ages $>30$ yr
(cfr. Fig.1, lower panel) are included for the montecarlo simulations.
Independently of the sample considered, initial periods $P_0\la 40$ ms
are inconsistent with the SN data.}
\end{figure}

Figure 2 shows the fraction $f_{sim}^c$ of the montecarlo simulations
of the SN sample for which the luminosity of each pulsar is found
below that of the corresponding SN. The sample of SNe selected is
either the one with ages $>10$ yr (top panel), or the one with ages $>
30$ yr (bottom panel), which allow tighter constraints while
minimizing optical depth effects.  Monte Carlo realizations of the
samples have been run for 50 Gaussian distributions of the period with
mean in the range $20-100\;$ms, and, for each of them, 4 values of
the dispersion\footnote{The dependences with $\sigma_{P_0}$ should be
taken as representative of the general trend, since it is likely that
$\sigma_{P_0}$ and $P_0$ might be correlated. But since no such
correlations have been studied and reported, we took as illustrative
example the simplest case of a constant $\sigma_{P_0}$ for a range of
$P_0$.}  $\sigma_{P_0}$ between 1 and $50\;$ms.  For each value of the
period, we performed 100,000 random realizations\footnote{The number
of random realizations is smaller here with respect to Fig.1 for
computational reasons since, while each panel of Fig.1 displays a
Monte Carlo simulation for one set of parameters only, each panel of
Fig. 2 is the results of 200 different Monte Carlo realizations. For a
few cases, however, we verified that the results were statistically
consistent with those obtained with a larger number of random
realizations.}.  Details of the results vary a bit between the
two age-limited subsamples.  This is not surprising since the extent
to which we can draw limits on the pulsar periods depends on the
measurements/limits of the X-ray luminosities of the SNe in the
sample. In a large fraction of the cases, we have only upper limits,
and therefore our analysis is dependent on the sensitivity at which
each object has been observed.  Independently of the sample, however,
we find that, for initial periods $P_0\la 35-40$ ms, the distribution
of pulsar luminosities is highly inconsistent with the SN data for any
value of the assumed width of the period distribution.

We need to emphasize that the specific value of $f_{sim}^c$ as a
function of $P_0$ should be taken as representative. Various authors
have come up with slightly different fits for the $L_x-\dot{E}_{\rm
rot}$ correlation.  If, for example, instead of the fit by Possenti et
al. (2002) we had used the fit derived by Guseinov et al. (2004), then
the limits on the period would have been more stringent.  On the
other hand, if, for some reason, the efficiency $\eta_x$ of conversion
of $\dot{E}_{\rm rot}$ into $L_x$ becomes low at high $\dot{E}_{\rm
rot}$, then our results would be less constraining. In order to assess
the robustness of our results with respect to changes in $\eta_x$, we
ran simulations of the pulsar population assuming that, for
$\dot{E}_{\rm rot}>\dot{E}_{\rm rot} ^{\rm max, obs}\sim 10^{39}$ erg
s$^{-1}$ (where $\dot{E}_{\rm rot} ^{\rm max, obs}$ is the maximum
observed $\dot{E}_{\rm rot}$), the efficiency becomes
$\eta_x^{\prime}=\epsilon \eta_x$, and we tried with a range of values of
$\epsilon <1$. This test also addresses the issue of a bias in our
results deriving from a possible shallower slope for the youngest
pulsars of the population, as discussed in \S3. We ran Monte Carlo
simulations for the ACC birth parameters, and decreased $\epsilon$ by
increments of 0.02. We found that, only for the very low value
$\epsilon\sim 10^{-4}$, a sizable fraction of $\sim 5\%$ of the
simulations predicts pulsar X-ray luminosities that are fully consistent
with the SN data. Therefore, we conclude that our results on the millisecond
birth periods of pulsars are reasonably robust with respect to uncertainty
in the $L_x-\dot{E}_{\rm rot}$ for the youngest members of the
population.

Another systematic that might in principle affect our results would arise
if a fraction of neutron stars is born with a non-active
magnetosphere so that their X-ray luminosity at high $\dot{E}_{\rm
rot}$ is much smaller than for the active pulsars, then the limits on
the initial periods of the ``active'' pulsars would be less stringent.
An example of non-active neutron stars could be that of the CCOs
discussed in \S3. However, until the fraction of these stars becomes
well constrained by the observations and an independent
$L_x-\dot{E}_{\rm rot}$ is established for them, it is not possible to
include them quantitatively in our population studies.  Similarly, the
precise fraction of BHs versus NSs in the remnant population plays a
role in our results. A larger fraction of BHs would alleviate our
constraints on the initial spin periods, while a smaller fraction
would, obviously, make them tighter. If a fraction of those BHs had a
luminosity larger than the maximum assumed upper limit in our
simulations (due to e.g. accretion from a fallback disk as discussed
above), then our results would again be more constraining.  While our
work is the first of its kind in performing the type of analysis that
we present, future studies will be able to improve upon our results,
once the possible systematics discussed above are better constrained,
and deeper limits are available for the full SN sample.

\section{Summary}

In this paper we have proposed a new method for probing the birth
parameters and energetics of young neutron stars.  The idea is simply
based on the fact that the X-ray measurements of young supernovae
provide upper limits to the luminosity of the young pulsars embedded
in them. The pulsar X-ray luminosity on the other hand, being directly
related to its energy loss, provides a direct probe of the pulsar spin
and magnetic field.  Whereas pulsar birth parameters are generally
inferred through the properties of the radio population, the X-ray
properties of the youngest members of the population provide tight and
independent constraints on those birth parameters, and, as we
discussed, especially on the spins.

The statistical comparison between theoretical predictions and the
distribution of X-ray luminosity limits that we have performed in this
work has demonstrated that the two are highly inconsistent if the bulk
of pulsars is born with periods in the millisecond range.  Whereas we
cannot exclude that the efficiency $\eta_x$ of conversion of
rotational energy into X-ray luminosity could have a turnover and drop
at high values of $\dot{E}_{\rm rot}$ to become $\eta_x\ll 1$, the
2-10 keV pulsar data in the currently observed range of $\dot{E}_{\rm
rot}$ do not point in this direction (but rather point to an increase
of $\eta_x$ with $\dot{E}_{\rm rot}$), and there is no theoretical
reason for hypothesizing such a turn over.  However, even if such
a turnover were to exist just above the observed range of
$\dot{E}_{\rm rot}$, we found that only by taking an efficiency
$\eta_x^{\prime}\sim 10^{-4}\eta_x$ above $\dot{E}_{\rm rot}^{\rm
max,obs}$, our results would lose their constraining value for the ms
spin birth distributions. Therefore, we can robustly interpret our
results as an indication that there must be a sizable fraction of
pulsars born with spin periods slower than what has been derived by a
number of radio population studies as well as by hydrodynamic
simulations of SN core-collapse (e.g. Ott et al 2006).  Our findings
go along the lines of a few direct measurements of initial periods of
pulsars in SNRs (see e.g. Table 2 in Migliazzo et al. 2002), as well
as some other population synthesis studies (Faucher-Gigure \& Kaspi
2006; Ferrario \& Wickramasinghe 2006; Lorimer et al. 2006).  Our
results for the bulk of the pulsar population, however, do not exclude
that the subpopulation of magnetars could be born with very fast
spins, as needed in order to create the dynamo action responsible for
the $B$-field amplification required in these objects (Thompson \&
Duncan 1993).  Because of their very short spin-down times, the energy
output of magnetars can be dominated by the spin down luminosity only
up to timescales of a fraction of year, during which the SN is still
too optically thick to let the pulsar luminosity go
through. Therefore, our analysis cannot place meaningful constraints
on this class of objects.

Finally, our results also bear implications on the contribution of
young pulsars to the population of the Ultra Luminous X-ray sources
(ULXs) observed in nearby galaxies. The model in \S3 predicts that a
sizable fraction of that population could indeed be made up of young,
fast rotating pulsars (Perna \& Stella 2004).  However, the analysis
performed here shows that the contribution from this component,
although expected from the tail of the population, cannot be as large
as current models predict.
 
The extent to which we could perform our current analysis has been
limited by the size of the SNe sample, and, especially, by the
available X-ray measurements.  The fact that, in a large fraction of
the sample, we have limits rather than detections, means that a large
improvement can be made with deeper limits from longer observations.
The deeper the limits, the tighter the constraints that can be derived
on the spin period distribution of the pulsars.  The analysis proposed
and performed here is completely uncorrelated from what done in radio
studies, and therefore it provides an independent and complementary
probe of the pulsar spin distribution at birth (or shortly
thereafter); our results provide stronger constraints on theoretical
models of stellar core collapse and early neutron star evolution,
making it even more necessary to explain why neutron stars spin down
so rapidly immediately after birth (see also Thompson, Chang \&
Quataert 2004; Metzger, Thompson \& Quataert 2007).

\section*{Acknowledgements}

We thank Roger Chevalier, John Raymond and Stefan Immler for very
useful discussions on several aspects of this project. We are
especially grateful to Bryan Gaensler and Shami Chatterjee for their
careful reading of our manuscript and detailed comments.  We also
thank the referee for his/her insightful suggestions.  RP and RS thank
the University of Sydney for the partial support and the kind
hospitality during the initial phase of this project.

{}

\end{document}